\documentclass[12pt]{article}

\catcode`\@=11
\@addtoreset{equation}{section}
\catcode`\@=12

\newcommand{\bea}{\begin{eqnarray}}
\newcommand{\eea}{\end{eqnarray}}
\newcommand{\hp}{\hat{\phi}}
\newcommand{\TPM}{'t Hooft-Polyakov monopole}
\def\be{\begin{equation}}
\def\ee{\end{equation}}

\def\fr{\frac}
\def\a{\alpha}
\def\b{\beta}

\def\d{\delta}
\def\e{\epsilon}

\def\l{\lambda}

\def\x{\xi}
\def\th{\theta}
\def\w{\omega}

\def\d{\delta}

\def\L{\Lambda}

\def\p{\partial}
\def\t{\tilde}

\def\nn{\noindent}
\def\no{\nonumber}

\def\cS{{\cal S}}

\title{{\small\hfill IMSc/2002/12/41}\\ \textbf{Topological field
    patterns of the Yang-Mills theory}}

\author{E. Harikumar\footnote{hari@imsc.res.in},~
Indrajit Mitra\footnote{indrajit@imsc.res.in}~ and
H. S. Sharatchandra\footnote{sharat@imsc.res.in} \\\\
The Institute of Mathematical Sciences,\\ C.I.T. Campus, Taramani P.O.,\\
Chennai 600 113, India}
\date{}

\begin{document}
\maketitle

\begin{abstract}

It is shown that the $SO(3)$ gauge
field configurations can be completely characterised by certain gauge
invariant vector fields.  The singularities of these vector fields
describe the topological aspects of the gauge field configurations.
The topological (or monopole) charge is expressed in terms 
of an Abelian vector potential.

\end{abstract}
\nn Keywords: Monopole, Abelianisation, Poincar\'e-Hopf Index.\\
\nn PACS no: 14.80.Hv, 11.15.-q, 11.15.Tk\\
\newpage
%%%%%%%%%%%%%%%%%%%%%%%%%
\section{Introduction}
%\label{s:intro}
%%%%%%%%%%%%%%%%%%%%%%%%%
In this letter we develop a formalism to describe generic
topological field patterns of the Yang-Mills theory. This is done
using certain gauge invariant special directions (actually orthonormal
frames) \cite{rps} provided by the non-Abelian magnetic fields. These
are the analogues of the Ricci principal directions \cite{rps} in
general relativity. The singularities of these frames locate the
topological aspects. We illustrate this explicitly in the case of the
\TPM~\cite{tpm}. This formalism provides a characterisation of the
\TPM~using only the gauge fields, even in the interior. The framework
is in the spirit of the Abelian projection procedure of 't Hooft
\cite{th}. The procedure has some connection with Ref.~\cite{manu}, but
our emphasis is to use only the gauge fields and not the Higgs.

The topological character of the \TPM ~\cite{tpm} is mainly described
by the behaviour of the Higgs field $\phi^a(x)$, $a=1,2,3$, at spatial
infinity: $\phi^a(x)\sim x^a/r$ for large $r$($=\sqrt{x^ax^a}$). It
has been recognized from the beginning \cite{arafune} that such a
behaviour of the Higgs field requires it to be zero at some point
$x_0$ in the interior which may be identified with the `centre' of the
\TPM. Using the Higgs field, we can construct the Poincar\'e -Hopf
current \cite{arafune}
\bea
k_i=\frac{1}{2}\epsilon_{ijk}\epsilon_{abc}\hp^a\partial_j\hp^b\partial_k\hp^c
\eea 
where $\hp^a=\phi^a/(\phi^b\phi^b)^{1/2}$ is the normalized Higgs
field. This is divergenceless except at the centre $x_0$ where $\hp^a$
is undefined:
\bea 
\partial_i k_i(x)=4\pi\delta^{(3)}(x-x_0).  
\eea

The monopole charge is thereby related to the topological charge, 

\bea
M&=&\frac{1}{4\pi}\int d^3x\,\partial_i k_i(x)\nonumber\\
&=&\frac{1}{4\pi}\oint_S dS^i k_i(x), 
\eea 
where the surface integration can be carried over any surface $S$
enclosing the centre.  This counts the number of times the normalized
Higgs field covers the unit sphere in the isospace when we cover the
surface $S$ once. In this way the Poincare-Hopf index of the isolated
zero of the Higgs field gives the topological character of the \TPM.

The \TPM~may also be described by gauge field using the 't Hooft
Abelian magnetic field \cite{tpm}
\bea
b_i=\hp^a B{_i^a} +\frac{1}{2}\epsilon_{ijk}\epsilon_{abc}\hp^aD_j\hp^bD_k\hp^c
\eea
where
$B{_i^a}=\epsilon_{ijk}(\partial_jA{_k^a}-\frac{1}{2}\epsilon_{abc}
A{_j^b}A{_k^c})$ is the non-Abelian magnetic field and
$D_i\hp^a=\partial_i\hp^a -\epsilon_{abc}A{_i^b} \hp^c$ is the
covariant derivative of the (normalized) Higgs field. We have
\bea
b_i=\epsilon_{ijk}\partial_j a_k +k_i,                                         
\label{eq:5}
\eea
where $a_i=\hp^a A{_i^a}$. The magnetic charge $\partial_i b_i$ comes
entirely from the topological charge defined above: $\partial_i
b_i(x)=4\pi\delta^{(3)}(x-x_0)$.

%%%%%%%%%%%%%%%%%%%%%%%%%%%%%%%%%%%%%%%%%%%%%%%%%%%%%%%%%%%%%%%%%%%%%%%%
\section{Topological characterisation using only gauge fields}
%%%%%%%%%%%%%%%%%%%%%%%%%%%%%%%%%%%%%%%%%%%%%%%%%%%%%%%%%%%%%%%%%%%%%%%%
  It is to be noted that the above characterisation using the non-Abelian
  gauge potential, necessarily uses the Higgs field also. In the
  asymptotic region, $r\to\infty$, generalised Stokes' theorem can be
  used to characterise the monopole using only the non-Abelian gauge
  field \cite{goddard,x}.  But this approach does not work in the
  interior region. A topological characterisation of monopole using only
  gauge field everywhere including the interior has been realised
  recently \cite{rps}. 
  
  A gauge invariant characterisation of monopole using only the
  non-Abelian gauge field is as follows: Consider a $3\times 3$, real,
  symmetric matrix 
  \be S_{ij}(x) =B{_i^a}(x)B_{j}^a(x)
  \label{sij} 
  \ee
  which is gauge invariant. Consider the eigenvalue equation 
  \be
  S_{ij}(x)\zeta_{j}^A(x)=\l^A(x)\zeta_{i}^A(x),~~A=1,2,3.
  \label{gieve}  
  \ee 
  Here there is no summation over $A$. Since $S_{ij}$ is symmetric,
  the normalised eigenvectors $\zeta_{i}^A(x)$ provide an orthonormal
  frame at each $x$, i.e.,
  \be
  \zeta{_i^A}(x)\zeta_{i}^B(x)=\d^{AB}.
  \label{frame}
  \ee 
  This frame is invariant under the local gauge transformations.  The
  topological aspects of the monopole configuration can be related to
  the singularities of these three vector fields $\zeta_{i}^A(x),
  A=1,2,3.$ To be specific, one of these, say $\zeta_{i}^1$, will
  appear to be diverging from the centre of the monopole. Thus here we
  have three orthonormal vectors to characterise the topological
  aspects completely in contrast with one unit vector used in the
  Faddeev-Niemi ansatz \cite{fn}.
  
  It is to be noted that the set $\l^A(x)$ and $\zeta_{i}^A(x)$
  together provide the complete gauge invariant information about the
  non-Abelian gauge potential $A_{i}^a$. Clearly $A_{i}^a$ ( or
  $B_{i}^a$) have nine degrees of freedom at each $x$, of which three
  are gauge degrees. The six gauge invariant degrees can be now
  described by the three fields $\l^A(x)$ and the $3\times 3$
  orthogonal matrix $(\zeta(x))_{iA}=\zeta_{i}^A(x)$ (this matrix has
  three degrees of freedom, e.g.,the Euler angles).

  Instead of using the eigenvectors of $S_{ij}$ which is quadratic in
  non-Abelian magnetic field, we may construct the frame $\zeta_{i}^A$
  as follows. The $3\times 3$ matrix $(B)_{ia}=B_{i}^a$ can be made
  symmetric at each $x$ by an appropriate local gauge transformation.
  This is because any real matrix can be expressed as $B_{i}^a=({\cS}
  O)_{ia}$, where $\cS$ is a real symmetric matrix ( not to be
  confused with $S_{ij}$) and $O$ is an orthogonal matrix. Here, $O$
  can be removed by the local gauge transformation $B_{i}^a\to
  O_{ab}B_{i}^b $. In this special gauge, $S_{ij}=(B^2)_{ij}$, and so
  $\zeta$'s are the eigenfunctions of $B_{i}^a$.
  
  Yet another way of constructing the frames $\zeta{_i^A}$ is as
  follows. By an appropriate local gauge transformation, we may make
  the three columns of the matrix $B_{i}^a$ mutually orthogonal (but
  not normalised). After normalisation, these columns give the frames
  $\zeta_{i}^A$.  The reason is that any $B_{i}^a= (O_{1}^T\L
  O_{2})_{ia}$, where $O_{1}$ and $O_{2}$ are orthogonal matrices and
  $\L$ is a diagonal matrix. By gauging away $O_{2}$, we get $B_{i}^a=
  (O_{1}^T\L )_{ia}$ (which has mutually orthogonal columns).
  Substitution into (\ref{sij}) then shows that $O_1 S O_1^T$ is
  diagonal, implying $(O_1^T)_{iA}=\zeta_i^A$. By normalizing the
  columns of $O_1^T\L$, one obtains precisely this matrix $O_1^T$.
  
  For the subsequent analysis, we find it more useful to consider the
  symmetric tensor
  \be
  S^{ab}(x)=B_{i}^a(x)B_{i}^b(x)\,,
  \label{gcovs}
  \ee 
  which is gauge covariant, instead of the gauge invariant tensor
  $S_{ij}$, and the normalised eigenfunctions $\xi_{a}^A$:
  \bea
  S^{ab}(x)\xi_b^A(x)=\lambda^A(x)\xi_a^A(x),~~ A=1,2,3.
  \eea
  It can readily checked that the eigenvalues are indeed the same for
  both the tensors, while the eigenfunctions are related:
  $B_{i}^a\xi_{a}^A$ is same as $\zeta_{i}^A$ up to a normalisation.
  (For Yang-Mills field configurations, generically the $3\times 3$
   matrix $B_{i}^a(x)$ is invertible \cite{ps}.)
  
  For each of $A=1,2$ and $3$, $\xi_{a}^A$ which is constructed from
  the non-Abelian gauge field, provides an isotriplet scalar, like the
  (normalised) Higgs field of the \TPM. We will use these Higgs like
  fields to characterise the topological aspects of the non-Abelian
  gauge fields. We first illustrate this for \TPM~for which $S^{ab}$
  has the general form
  \be
  S^{ab}=\a(r)\d^{ab} +\b(r)x^a x^b
  \label{tmps}
  \ee 
  where $\a$ and $\b$ are functions of only the radial distance $r$.
  One of the eigenvectors is the radial vector,
  $\xi_{a}^1=x^a/r$. The other two can be chosen to be any
  linearly independent combination of the basis vectors $\hat{\th}$
  and $\hp$ of the spherical coordinate system, and these two are
  degenerate eigenfunctions of $S^{ab}$. This double degeneracy is a
  consequence of the spherical symmetry of the \TPM~solution.
  
  It is important to note that the singularities in the eigenvector
  fields appear only at points where the eigenvalues become degenerate
  (because only at those points, the direction of an eigenvector
  can be indeterminate) \cite{th}.
  For example, in the case of \TPM, the eigenvalues are triply
  degenerate at the origin. Such a property is necessary because the
  entries of the matrix $B_{i}^a$ or $S^{ab}$ are themselves not
  singular at the origin. This provides a way to define centres of the
  monopoles (and other topological objects) for an arbitrary
  Yang-Mills field configuration.

 %%%%%%%%%%%%%%%%%%%%%%%%%%%%%%%%%%%%%%%%%%%%%%%%%%%%%%%%%%%%%%%%%%%%%%
 \section{Abelian vector potential for Poincar\'e - Hopf current}
 %%%%%%%%%%%%%%%%%%%%%%%%%%%%%%%%%%%%%%%%%%%%%%%%%%%%%%%%%%%%%%%%%%%%%%
  The eigenvector $\xi_{a}^1$ has unit topological charge at the
  origin. We construct the Poincar\'e -Hopf current for each of the
  three vectors $\xi_{a}^A$:
  \be
  k_{i}^A=\fr{1}{2}\e_{ijk}\e^{abc}\xi_{a}^A\p_j\xi_{b}^A\p_k\xi_{c}^A,
  \label{wni}
  \ee 
  where there is no summation over $A$. Since $\p_ik_i^A=0$ (except
  perhaps for Dirac delta function contribution due to the
  singularities of $\xi_{a}^A(x)$) we can express $k_{i}^A$ as a curl
  of a vector potential. We now obtain a formal expression for this
  vector potential. Regarding the orthogonal matrix
  $(\xi)_{aA}=\xi{_a^A}$ as a local gauge transformation, we get
  the corresponding pure gauge potential as
  \be
  \w_{i}^A=\fr{1}{2}\e^{ABC}\x_{a}^B\p_i\xi_{a}^C.
  \label{pureg}
  \ee 
  Using this $\w_{i}^A$, we re-express $k_{i}^A$ in terms of $\w_{i}^A$ as
  \bea
  k_{i}^A &=&\fr{1}{2}\e_{ijk}\e^{ABC}\xi_{b}^B\xi_{c}^C\p_j\xi_{b}^A\p_k\xi_{c}^A
                         ~~~~~ ({\rm no ~summation ~over ~A})\no\\
          &=&\fr{1}{2}\e_{ijk}\e^{ABC}\w_{j}^B\w_{k}^C,
  \label{kinpg}
  \eea 
  where we have used the fact $\det \xi_{aA}=1$ in the first step and
  $\e^{ABC}\e^{ABD}\e^{ACE}=\e^{ADE}$ (no sum over $A$) in the second. Since
  $\w_{i}^A$ is a pure gauge, the corresponding non-Abelian magnetic field
  vanishes, i.e., 
  \be \e_{ijk}\left
    (\p_j\w_{k}^A-\fr{1}{2}\e^{ABC}\w_{j}^B\w_{k}^C\right)=0.
  \label{pmag}
  \ee
  This allows us to write $k_{i}^A$ as a curl:
  \be
 k_{i}^A=\e_{ijk}\p_j\w_{k}^A.                          \label{curlcurl}
  \ee
  When monopole and other topological objects are present, some
  $\xi_{a}^A$ have singularities at the centres. Then, in general,
  $\p_ik_i^A$ will have Dirac delta function singularities. For such a
  situation (\ref{curlcurl}) is to be modified exactly in the same
  manner as Dirac's construction of the vector potential of a
  monopole:
  \be 
  k_{i}^A=\e_{ijk}\p_j\w_{k}^A ~- {\rm ~Dirac ~string~contributions~}.
  \label{kpcurl}
  \ee
  In this case, $\w_{i}^A$ is not strictly a pure gauge, and (\ref{pmag})
  gets modified to 
  \bea
  \e_{ijk}\left
      (\p_j\w_{k}^A-\fr{1}{2}\e^{ABC}\w_{j}^B\w_{k}^C\right)
      ={\rm ~Dirac ~string~contributions~}.                    \label{mmpmag}
  \eea

  By this procedure, we have succeeded in describing the topological
  objects of the non-Abelian gauge fields in an Abelian fashion. The
  topological features are contained in the (ordinary) curl of a
  vector potential ($\w_{i}^A$) without requiring the non-linear
  terms. We will now illustrate these features in the case of \TPM.

  The Poincar\'e-Hopf currents $k_{i}^A$ for the \TPM~are as follows.
  We have already taken $\xi_a^1={\hat x}^a$. Let us choose
  $\xi_a^2={\hat \theta}^a$ and $\xi_a^3={\hat \phi}^a$.
  On going over to spherical polar coordinates,
  (\ref{wni}) becomes
  \be
  k_{i}^A ={\hat x}_i\fr{1}{r^2 \sin\th}~\e^{abc}
           \xi_{a}^A\fr{\p\xi_{b}^A}{\p\th}\fr{\p\xi_{c}^A}{\p\phi}\,,
  \ee
  from which, for the \TPM, we get
  \bea 
   k_{i}^1&=&{\hat x}_i~ \fr{1}{r^2}~,\no\\
   k_{i}^2&=&{\hat x}_i~\fr{ \cot\th}{r^2}~,\no\\
   k_{i}^3&=&0.
   \label{kvalue}
  \eea 
  Here we note that $k_{i}^1$ is precisely the magnetic field of
  a Dirac monopole.  The flux over any surface enclosing the
  origin is then given by 
  \be 
  \oint dS^ik_{i}^1=4\pi.  
  \ee
  Note that the magnetic current corresponding to $A=2$, viz.
  $k_{i}^2$ is also non-zero. However, the corresponding magnetic
  charge is zero, i.e.,
  \be
  \oint dS^ik_{i}^2=0.\no
  \ee
  It corresponds to a radial flux from the region $z~<~0$ to the region $z~>~0$.
  
  Using (\ref{pureg}), we find the
  vector potentials $\w_{i}^A$ for the \TPM~are
  \bea
  \w_{i}^1&=&-{\hp}_i\fr{\cot\th}{r}\,,\label{w1}\\
  \w_{i}^2={\hp}_i\fr{1}{r}\,,&&\w_{i}^3=-{\hat\th}_i\fr{1}{r}\,.
  \label{wvalue} 
  \eea
  The potential $\w_{i}^1$ is to be compared with (appropriately scaled, viz.
  $A\to eA$) Dirac potential for a monopole 
  \be
  A_{i}(x)={\hp}_i\fr{eM}{4\pi r}\fr{(1-\cos\th)}{\sin\th}\,,
  \label{dpot}
  \ee 
  where $e$ is the electric charge and $M$ is the magnetic charge.
  This has the Dirac string along the negative $z-$axis. Consider the
  average of the Dirac potentials with string along the positive
  $z-$axis and positive $z-$axis (obtained from (\ref{dpot}) by
  changing $\th\to\pi+\th$). This gives the magnetic field of a
  monopole of Schwinger charge $M={4\pi}/{e}$ exactly same as the
  $\w_{i}^1$ in (\ref{w1}).  
  
  It can be explicitly checked that, on
  ignoring the Dirac string contribution, $\e_{ijk}\p_j\w_{k}^A$ gives
  precisely the same Poincar\'e~-Hopf current $k_{i}^A$, and the same
  winding numbers.  Thus, the vector potential for $A=2$ gives magnetic flux
  without monopole, while the vector potential for $A=3$ does not give
  magnetic flux. Their relevance is elucidated later.

  %%%%%%%%%%%%%%%%%%%%%%%%%%%%%%%%%%%%%%%%%%%%%%%%%%%%%%%%%%%%%%%%%%
  \section{`Abelianisation' of Yang-Mills potential}
  %%%%%%%%%%%%%%%%%%%%%%%%%%%%%%%%%%%%%%%%%%%%%%%%%%%%%%%%%%%%%%%%%
  Consider a gauge transformation using the orthogonal matrix
  $(\xi)_{aA}=\x_{a}^A$. The transformed potentials are
  \bea
  a_{i}^A &=& \xi_{a}^A A_{i}^a +\fr{1}{2}\e^{ABC}\xi_{a}^B\p_i\xi_{a}^C\no\\
  &=&{\t A}_{i}^A+\w_{i}^A.
  \label{gtym}
  \eea
  Here ${\t A}_{i}^A$ is the Abelian vector potential of 't Hooft now
  constructed using the three Higgs like fields $\xi_{a}^A$ ( compare
  $a_i$ given just after (\ref{eq:5})). When monopoles are present,
  this gauge transformation is singular due to singularities 
  of $\x_{a}^A$ (see (\ref{mmpmag})). As an
  explicit example we consider the \TPM~\cite{tpm}:
  \bea
  A_i^a=-\epsilon_{aij}{\hat x}^j\frac{1-K(r)}{r}\,,
  \eea
  where $K(r)\to 0$ as $r\to\infty$ and $K(r)=1 + O(r^2)$ as $r\to 0$. So we get
  \bea
  {\t A}_{i}^1&=&0,\no\\
  {\t A}_{i}^2 &=& -{\hp}_i \fr{1-K(r)}{r}\,,\no\\
  {\t A}_{i}^3&=&{\hat\th}_i\fr{1-K(r)}{r}\,.
  \label{AA}
  \eea
  Making use of (\ref{w1}) and (\ref{wvalue}), we then obtain
  \bea
  a_{i}^1&=&-{\hp}_i\fr{\cot\th}{r}\,,\no\\
  a_{i}^2&=& {\hp}_i\fr{K(r)}{r}\,,\no\\
  a_{i}^3&=&-{\hat\th}_i\fr{K(r)}{r}\,.
  \label{avalue}
  \eea 
  For $A=1$ we recover the Dirac potential of a point monopole
  ($a_{i}^2$ and $a_{i}^3$ vanish asymptotically) while the orthogonal
  transformation provided by $\xi_{a}^A$ rotates ${\hat x}^a$ into
  $1-$direction. This is similar to the singular gauge transformation
  rotating the Higgs field $\phi^a$ to the $3-$direction in
  Ref.~\cite{arafune}.  However, we obtained this transformation using only
  the gauge potential.

The relevance of $a_i^2$ and $a_i^3$ is the following.
In an Abelian theory, the energy of the Dirac monopole would diverge
due to the singularity of $a_i^1$ at the origin. Now however
$a_i^2$ and $a_i^3$ also diverge in a specific way to ensure that
the nonlinear terms $\epsilon_{ijk}a{_j^2}a{_k^3}$ in $B_i^1$ cancel the
singular contribution of  $a_i^1$.
  
  The three new potentials $a_{i}^A$, when regarded as three Abelian
  vector potentials, carry all the information of the non-Abelian
  topology. For $A=1$ we have a Dirac monopole charge, while the $A=2,3$
  cases have none.
  
  Note that $a_i^A=\xi_a^A(A_i^a-{\tilde\w}_i^a)$ where
  $\t\w_{i}^a=\fr{1}{2}\e_{abc}\x_{b}^B\p_i\xi_{c}^B$.
  Both $A^i_a$ and
 ${\tilde\w}_i^a$ transform inhomogeneously as gauge potentials
  under local gauge transformation. Therefore
  $A_i^a-{\tilde\w}_i^a$ transforms homogeneously as a triplet, and its
  scalar product with $\xi_a^A$ is invariant. Thus
  $a_{i}^A$ are gauge invariant under the non-Abelian gauge
  transformation acting on the subscript $a$ in (\ref{gtym}). The
  superscript $A$ provides gauge invariant directions as obtained from
  the eigenfunctions of $S_{ij}$. Instead of the non-Abelian Wilson
  loop ${\rm Tr}~{\cal P}\exp(i\oint dx^i A_i)$, we can consider three
  Abelian Wilson loops
\be
W^A[C]=\exp(i{\oint}_C dx^ia_{i}^A).
\ee
These are gauge invariant (under the non-Abelian gauge transformation)
and carry all topological information about the non-Abelian gauge
fields.

%%%%%%%%%%%%%%%%%%%%%%%%%%%%%%%%%%%%%%%%%%%%%%%%%%%%%%%%%%%%%%%%%%%%
\section{Conclusion}
%%%%%%%%%%%%%%%%%%%%%%%%%%%%%%%%%%%%%%%%%%%%%%%%%%%%%%%%%%%%%%%%%%%

In this letter, we have shown that the non-Abelian ($SO(3)$) gauge
field configurations can be completely characterised by certain gauge
invariant vector fields.  The singularities of these vector fields
describe the topological aspects of the gauge field configurations.
Our procedure provides an Abelianisation of the non-Abelian gauge
theory in two ways:
\begin{enumerate}
\item{The topological (or monopole) charge is characterised by the curl 
of an Abelian vector potential.}
\item{The non-Abelian gauge field is transformed to three `gauge invariant' 
vector potentials and they capture the topological aspects when treated as Abelian 
vector potentials.}
\end{enumerate}
Using this approach, we can obtain the most general topological field
patterns of the Yang-Mills fields. For example, one finds that the
generic configuration has half the \TPM~charge. Also there are
vortices of half-integral winding number. These aspects will be
elaborated elsewhere \cite{his}.

%\newpage

\end{document}